\documentclass[aps,prb,twocolumn,showpacs]{revtex4-1}
\usepackage{graphicx}
\usepackage{latexsym}
\usepackage{amssymb}
\usepackage{amsmath}
\usepackage{amsfonts}
\usepackage{bm}
\usepackage{multirow}
\usepackage{color}

\begin{document}

\title{Order parameter for multi-channel Kondo model at quantum criticality}

\author{Ru Zheng}
\email{zhengru@ruc.edu.cn}
\address{Department of Physics, Renmin University of China, Beijing 100872, China}
\author{Rong-Qiang He}
\email{rqhe@ruc.edu.cn}
\address{Department of Physics, Renmin University of China, Beijing 100872, China}
\author{Zhong-Yi Lu}
\email{zlu@ruc.edu.cn}
\address{Department of Physics, Renmin University of China, Beijing 100872, China}

\begin{abstract}
A multi-channel Kondo model, where two or more equivalent but independent channels of electrons compete to screen a spin-1/2 impurity, shows overcompensation of the impurity spin, leading to the non-Fermi liquid behavior in various thermodynamic and transport properties. However, when the channel symmetry is broken, an impurity quantum phase transition can occur at zero
temperature. Identification of an order parameter describing the impurity quantum phase transition is very difficult since it is beyond the conventional Landau-Ginzburg-Wilson theory. By employing the natural orbitals renormalization group method, we have studied both two-channel and three-channel Kondo models, from the perspective of spin correlation between the impurity and electrons in electronic channels. Here we demonstrate that by introducing spin correlation ratio as an order parameter we can characterize impurity quantum phase transitions driven by channel asymmetry. Especially the universal critical exponents $\beta$ of spin correlation ratio and $\nu$ of correlation length are explicitly determined by finite-size scaling analysis, namely $\beta = 0.10(1), \nu = 2.0(1)$ and $\beta = 0.10(1), \nu = 2.5(1)$ for the two-channel and three-channel Kondo models, respectively.
\end{abstract}


\maketitle

\section{Introduction}

The Kondo effect \cite{Kondo1964,Hewson1997}, induced by the antiferromagnetic exchange interaction between a localized spin-1/2 impurity and conduction electrons in the Fermi sea, is one of the intensively studied problems in quantum many-body physics. Below a characteristic energy scale $T_K$, namely the Kondo temperature, the magnetic impurity is collectively screened by the surrounding conduction electrons, leading to the Fermi liquid behavior in low-temperature properties. The concise interpretation of the Kondo effect is through the well-known single-impurity Kondo model~\cite{Kondo1964,Wilson1975,Bulla2008}, whose ground state is a Kondo singlet. On the other hand, the Kondo screening is strongly modified if the magnetic impurity is coupled to two or more independent channels of electrons.

Nozi\`eres and Blandin~\cite{Blandin1980} proposed a multi-channel generalization of the standard Kondo model, i.e., the multi-channel Kondo (MCK) model, where $M>1$ equivalent but independent channels of electrons compete to screen a spin-1/2 impurity and then the impurity is ultimately overscreened. Correspondingly, the various physical properties have been studied by Bethe ansatz~\cite{Andrei1984,Tsvelick1984,Tsvelick1985,Andrei1995,Andrei2002}, conformal field theory (CFT)~\cite{Affleck1991PRL,Affleck1991,Affleck1992,Affleck1993,Parcollet1998}, bosonization~\cite{Emery1992,Clarke1993} and NRG\cite{Pang1991,Affleck1992,Bulla2008}, as well as other methods~\cite{Sengupta1994,Fabrizio1995}. At low temperatures, the spin-1/2 impurity is simultaneously screened by each channel, resulting in the non-Fermi liquid behavior in various thermodynamic and transport properties. These nontrivial low-temperature properties include the non-vanishing zero-temperature impurity entropy $S_{\text{imp}}(T=0)=\ln [2\cos (\frac{\pi}{2+M})]$ and the fractional power-law behavior of impurity magnetic susceptibility $\chi_{\text{imp}}\propto T^{2\Delta-1}$ and of impurity specific heat ratio $\gamma_{\text{imp}}=C_{\text{imp}}/T \propto T^{2\Delta-1}$  for $T\to 0$, where $1 + \Delta = 1 + 2/(2+M)$ is the scaling dimension of the leading irrelevant operator at the fixed point. In addition, the resistivity at low temperature goes as $R \propto T^\Delta$, which is different from that of the Fermi liquid with $R \propto T^2$. In particular, in the two-channel case $M = 2$, i.e., the two-channel Kondo (2CK) model, logarithmic corrections in the thermodynamic properties occur instead, leading to $\chi_{\text{imp}}\propto \ln T$ and $\gamma_{\text{imp}} \propto \ln T$ for $T\to 0$ with an anomalous Wilson ratio $R_W = \chi_{\text{imp}}/\gamma_{\text{imp}}= 8/3$, in contrast to the result of the standard single-impurity Kondo model $R_W=8/4=2$. On the experimental side, several realizations of the 2CK effect have also been obtained~\cite{Cichorek2005,Potok2007, Mebrahtu2013,Keller2015,Iftikhar2015,Zhu2015,Cichorek2016}.

The above anomalous low-temperature properties of an MCK system are based on the condition that all the channels are equivalent, namely symmetric. However, the non-Fermi liquid physics is extremely delicate against the channel symmetry breaking perturbation: even the smallest asymmetry destabilizes the non-Fermi liquid fixed point in the renormalization group (RG) flow. For instance, in the 2CK model with $J_a$ and $J_b$ denoting the two Kondo couplings, when the channel symmetry is broken ($J_a \neq J_b$), a new temperature scale $T^* \propto (J_a-J_b)^2$, along with a corresponding length scale $\xi^* \sim 1/T^*$, for the crossover from the unstable overscreened non-Fermi-liquid fixed point to the stable fully screened Fermi-liquid fixed point has been found~\cite{Pang1991,Mitchell2011}. In this case, $T^*$ characterizes the energy scale for flow away from the overscreened fixed point at intermediate temperatures and the impurity is hence completely screened by the channel with the stronger coupling at $T = 0$. Only on fine tuning the Kondo couplings $J_a \to J_b$ to the symmetry point with $T^* \to 0$, one obtains the non-Fermi liquid physics at the lowest energy scales. Therefore at $T = 0$, an MCK system may undergo an impurity quantum phase transition (iQPT)~\cite{Vojta2006} driven by channel asymmetry, where only the impurity contribution to the free energy becomes singular at the critical point. The simple physical picture is that one channel couples to the impurity more strongly than the rest, then the impurity is screened only by this channel while the other channels decouple, leading to the standard one-channel Kondo (1CK) physics with a Fermi liquid phase~\cite{Blandin1980,Fabrizio1995,Mitchell2014}. An iQPT can then occur at the channel-symmetric point as the channel asymmetry is varied.

To describe a phase transition, we need to introduce an order parameter. But interpretation for iQPTs goes beyond the classical Landau-Ginzburg-Wilson scenario due to the fact that an iQPT is distinct from any bulk phase transitions, thus identification of an order parameter describing an iQPT is very difficult. Recently, great efforts have been made to explore order parameters characterizing an iQPT~\cite{Bayat2014,Alkurtass2016,Bayat2017}. Inspired by quantum information, Schmidt gap~\cite{Bayat2014} and negativity \cite{Alkurtass2016} have been introduced as order parameters to describe the iQPT in a 2CK model. Nevertheless, there is still lack of order parameters based on correlation functions, especially an observable one.

In this work, using the natural orbitals renormalization group (NORG) method~\cite{He2014}, we have explored such order parameters to describe the iQPTs driven by the channel asymmetry in both 2CK and three-channel Kondo (3CK) models, from the perspective of spin correlation between the impurity and electrons in the electronic channels. The corresponding universal critical exponents can be further extracted by finite-size scaling analysis.

This paper is organized as follows. In Sec.~\ref{sec:Model-Method}, the 2CK and 3CK models as well as the NORG numerical method are introduced. In Sec.~\ref{sec:correlation ratio}, we study the spin correlation between the impurity and electrons in electronic channels. The spin correlation ratio is then introduced and further demonstrated as an appropriate order parameter to describe the iQPTs driven by the channel asymmetry in both models. Finally, by finite-size scaling analysis, we explicitly determine the universal critical exponents $\beta$ of spin correlation ratio and $\nu$ of correlation length to be $\beta = 0.10(1), \nu = 2.0(1)$ and $\beta = 0.10(1), \nu = 2.5(1)$ for the 2CK and 3CK models, respectively. Section~\ref{sec:conclusion} gives a short discussion and summary of this work.

\section{Models and numerical method}
\label{sec:Model-Method}
\begin{figure}[htp!]
\centering
\includegraphics[width=0.9\columnwidth]{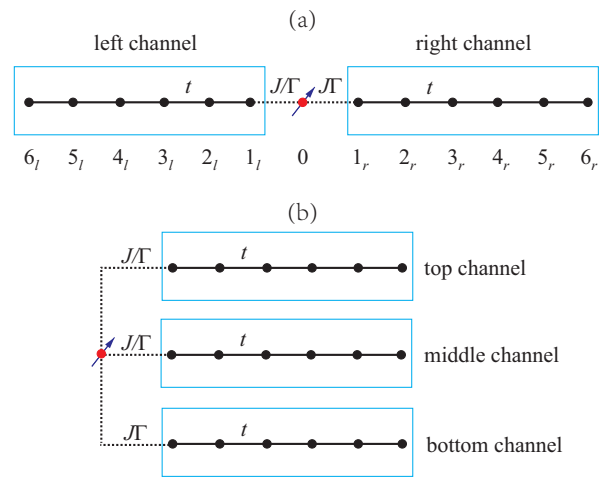}
\caption{\label{fig:Model}(color online) Schematics of (a) 2CK and (b) 3CK models with the electronic channels being simulated by independent periodic tight-binding chains. (a) A spin-1/2 impurity is coupled to its left and right channels by Kondo couplings $J_l=J/\Gamma$ and $J_r = J\Gamma$ with $\Gamma$ standing for the asymmetry parameter, respectively. For $\Gamma = 1$, the impurity is overscreened by both channels, resulting in the 2CK physics. While for $\Gamma \neq 1$ in the thermodynamic limit, the impurity is screened only by the channel with the stronger coupling, representing the 1CK physics. (b) A spin-1/2 impurity is coupled to the top, middle and bottom electronic channels with Kondo couplings $J_t=J_m=J/\Gamma$ and $J_b = J\Gamma$. The 3CK physics occurs at the symmetric point $\Gamma = 1$ with the impurity is screened simultaneously by all the channels. While for any $\Gamma < 1$, the 2CK physics emerges with the impurity being overscreened by the top and middle channels, while the bottom channel decouples. In contrast, for any $\Gamma > 1$, the 1CK physics emerges with the impurity being screened only by the bottom channel.}
\end{figure}

Generally, the Hamiltonian for an MCK model, where $M$ independent channels of electrons compete to screen a single spin-1/2 impurity, can be written as $H_{\text{MCK}} = \sum_a H_a + H_{\text{int}}$ with
\begin{equation}
\begin{array}{l}
H_a = \sum\limits_{k\sigma}\varepsilon_{k} c_{ak\sigma }^ {\dagger} c_{ak\sigma}, \\
H_{\text{int}}=\sum\limits_a J_a  {\bf{S}}_0 \cdot {\bf{s}}_{a1},
\end{array}
\label{eq:MCKM}
\end{equation}
where $a = 1,2,...M$ is the electronic channel index. The operator $c_{ak\sigma }^ {\dagger}$ creates an electron at a Bloch state with wavevector $k$ and spin $\sigma=\uparrow,\downarrow$ in channel $a$ with $\varepsilon_{k}$ denoting the dispersion relation. The impurity spin ${\bf{S}}_0$ is coupled to the electron spin ${\bf{s}}_{a1}$ at site 1 in channel $a$ via Kondo coupling $J_a$. Here we consider the tight-binding chain emulation of the MCK model, as shown in Fig.~\ref{fig:Model}, in which a localized spin-1/2 impurity is coupled to $M$ independent periodic tight-binding chains by antiferromagnetic exchange interaction.

A typical model for studying iQPTs in MCK system is the 2CK model, where there exists a nontrivial critical crossover at the channel-symmetric point with the impurity being overscreened by the two channels. As illustrated in Fig.~\ref{fig:Model}(a), the Hamiltonian for the 2CK model is given by $H_{\text{2CK}} = \sum_{a}H_a + H_{\text{int}}$ with
\begin{equation}
\begin{array}{l}
H_a = -t\sum\limits_{ij\sigma}(c_{ai\sigma }^ {\dagger}{c_{aj\sigma }}+ h.c.), \\
H_{\text{int}}=J/\Gamma{\bf{S}}_0 \cdot {\bf{s}}_{l1} + J\Gamma {\bf{S}}_0 \cdot {\bf{s}}_{r1}.
\end{array}
\label{eq:2CKM}
\end{equation}
Here, $c_{ai\sigma }^ {\dagger}({c_{ai\sigma }})$ represents the creation (annihilation) operator of a conduction electron at site $i$ in channel $a\in\left\{l,r\right\}$, ${\bf{s}}_{a1} =\frac{1}{2}\sum_{\alpha\beta}c_{a1\alpha}^ {\dagger}{{\bf{\sigma}}_{\alpha\beta}}{c_{a1\beta}}$ with ${\bf{\sigma}}$ representing the vector of Pauli matrices. The system size is $N = L + 1 = N_l + N_r + 1$ with $N_a$ denoting the number of sites in channel $a$ and $L$ the number of total sites in all the electronic channels, and we take $N_l = N_r$ in this work. We keep the Kondo couplings $J_l=J/\Gamma$ and $J_r = J\Gamma$, where the dimensionless quantity $\Gamma$ plays a controlling role. The system presents critical behavior around $\Gamma_c = 1$, where the two channels are symmetric with $J_l = J_r$, resulting in the 2CK physics with non-Fermi liquid phase. For any $\Gamma \neq 1$ in the thermodynamic limit $N \to \infty$, the 1CK physics with Fermi liquid phase emerges, and the impurity is screened only by the channel with the stronger coupling, while the other channel decouples. Hence, the $\Gamma = 1$ point acts as a critical point separating the two phases.

Another typical model we consider is the 3CK model. Likewise, the Hamiltonian for the 3CK model presented in Fig.~\ref{fig:Model}(b) can be given by $H_{\text{3CK}} = \sum_{a}H_a + H_{\text{int}}$ with
\begin{equation}
H_{\text{int}}=J/\Gamma{\bf{S}}_0 \cdot {\bf{s}}_{t1} + J/\Gamma{\bf{S}}_0 \cdot {\bf{s}}_{m1} + J\Gamma {\bf{S}}_0 \cdot {\bf{s}}_{b1}.
\label{eq:3CKM}
\end{equation}
Here, a spin-1/2 impurity is coupled to three (top($t$), middle($m$), and bottom($b$)) tight-binding chains with Kondo couplings $J_t = J_m = J/\Gamma$ and $J_b = J\Gamma$. Again, the system size is $N = L + 1 = \sum_{a} N_a + 1$ with $a\in\left\{t, m,b\right\}$ and $L$ denoting the number of total sites in all the electronic channels, and we take $N_t = N_m = N_b$ in this work. The 3CK physics with non-Fermi liquid phase occurs around the critical point $\Gamma = 1$, where the Kondo couplings to all the channels are equal, namely the three channels are symmetric, and thus the impurity is overscreened by all the channels. For any $\Gamma > 1$ in the thermodynamic limit $N \to \infty$, the impurity is screened only by the bottom channel, leading to the emergence of the standard 1CK physics. In contrast, for any $\Gamma < 1$, the 2CK physics emerges with the top and medium channels equally competing for screening the impurity, while the bottom channel decouples. As a result, the channel-symmetric 3CK point $\Gamma=1$ separates two distinct phases, namely the 2CK non-Fermi liquid phase for $\Gamma < 1$ and the 1CK Fermi liquid phase for $\Gamma > 1$. Recently, the global phase diagram of the three-channel spin-orbital Kondo model, which is relevant for Hund metals, has been studied using the NRG method~\cite{Wang2020}.

We adopted the NORG approach to study the critical behavior in the 2CK and 3CK models. The NORG method works efficiently on quantum impurity models in the whole coupling regime, and it preserves the whole geometric information of a lattice. \cite{He2014,He2015,Zheng2018}We emphasize that the effectiveness of the NORG is independent of any topological structure of a lattice. The NORG method works in the Hilbert space constructed from a set of natural orbitals~\cite{Lin2013,He2014}. Its realization essentially involves a representation transformation from site representation into natural orbitals representation through iterative orbital rotations. In practice, to efficiently realize the NORG, we only rotate the orbitals of bath, that is, only the bath orbitals are transformed into natural orbitals representation. By using the NORG method we can solve hundreds of non-interacting bath sites with any topological structures, while the computational cost is about $O(N_{\text {bath}}^3)$ with $N_{\text {bath}}$ being the number of bath sites.

Throughout the whole work, we took even number of sites in each channel, in which case the particle-hole symmetry is preserved. We set the nearest-neighbor hopping integral $t=1/2$ and kept half-filling of the conduction bands. All the calculations were carried out in the subspace of $S_{\rm total}^z=1/2$.

\section{Spin correlation ratio as an order parameter}
\label{sec:correlation ratio}
Screening of the single magnetic impurity in an MCK model by the electronic channels, no matter whether full screening as in the standard 1CK physics or overscreening as in the 2CK and 3CK physics, certainly manifests itself in the structure of the spin correlation between the localized impurity and electrons in the electronic channels in the ground state. Moreover, the spin correlation may be measured by spin-polarized scanning tunneling microscopy (STM). To be specific, the spin correlation between the impurity and electrons at site $i$ in channel $a$ has the following form of
\begin{equation}
C_a(i)=\langle 0| S_0^zs_{ai}^z|0\rangle,
\label{eq:correlation}
\end{equation}
where $|0\rangle$ denotes the ground state. We then consider the total correlation in channel $a$ as well as in all the channels, which are given by
\begin{equation}
\begin{array}{l}
C_a=\sum\limits_{i}C_a(i)=\sum\limits_{i}\langle 0| S_0^zs_{ai}^z|0\rangle, \\
C = \sum\limits_aC_a=\sum\limits_{ai}\langle 0| S_0^zs_{ai}^z|0\rangle.
\end{array}
\label{eq:correlation}
\end{equation}

\begin{figure}[htp!]
\centering
\includegraphics[width=0.8\columnwidth]{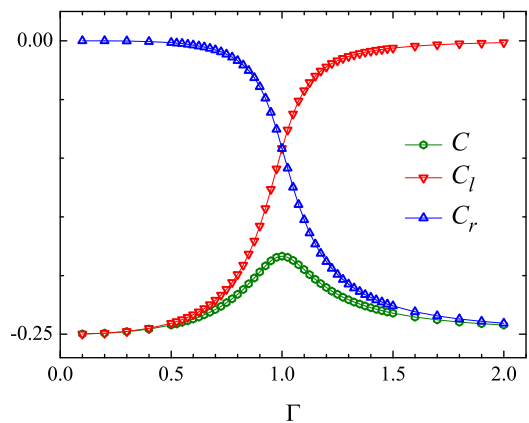}
\caption{\label{fig:Corrs}(color online) Ground state spin correlations $C$, $C_l$ and $C_r$ between the localized impurity and electrons in the electronic channels in the 2CK model, versus $\Gamma$ for system of size $N = 205$ with the fixed coupling $J = 1$.}
\end{figure}

We first take the 2CK model (Eq.~(\ref{eq:2CKM})) illustrated in Fig.~\ref{fig:Model}(a) into consideration. When the two channels are symmetric, namely $J_l=J_r$, the 2CK physics is valid with the impurity being overscreened by the two channels, where the temperature scale $T^* \propto (J_l-J_r)^2$ vanishes and the diverging length scale $\xi^* (\sim 1/T^*) \to \infty$ emerges in the thermodynamic limit. On the other hand, in the channel-asymmetric case, the standard 1CK physics emerges, where the magnetic impurity is screened only by the channel with the stronger coupling in the thermodynamic limit, while the other channel decouples. It is thus expected that the total correlation in a decoupled channel is 0, namely $C_a=0$ with $a$ representing the decoupled channel, meaning that there is no correlation between this channel and the impurity.

We have studied the 2CK model by using the NORG method. In Fig.~\ref{fig:Corrs}, we plot the calculated total correlations $C$, $C_l$ and $C_r$ as functions of $\Gamma$ with the fixed coupling $J = 1$. As the figure shows, when the parameter $\Gamma \ll 1$,  $C_r \to 0$ and $C_l \to -1/4$, indicating that the impurity is correlated only with the left channel. In contrast, the impurity is correlated only with the right channel for $\Gamma \gg 1$. $C_l$ and $C_r$ cross at the critical point $\Gamma =1$, where the 2CK physics is valid with the impurity being equally correlated with the two channels. It is expected that for any $\Gamma \ne 1$ in the thermodynamic limit, the correlation in channel $a$ with smaller coupling $C_a \to 0$ while $C_{\bar a} \to -1/4$, where $\bar a$ denotes the other channel with the stronger coupling. Deviations of $C_a$ and $C$ from the values in thermodynamic limit in Fig.~\ref{fig:Corrs} result from the finite-size effect of the system we choose.

\begin{figure}[htp!]
\centering
\includegraphics[width=1.0\columnwidth]{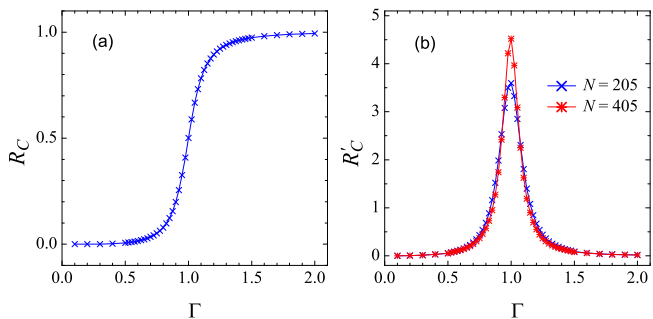}
\caption{\label{fig:CorrRatio}(color online) (a) Spin correlation ratio $R_C$ as a function of $\Gamma$ for system of size $N = 205$ and (b) derivative of $R_C$ with respect to $\Gamma$, namely $R_C^{\prime}=\partial R_C/{\partial\Gamma}$, for systems of size $N=205$ and $N = 405$ with the fixed coupling $J = 1$ in the 2CK model.}
\end{figure}

We hence introduce the spin correlation ratio $R_C$ by the following form of
\begin{equation}
R_C= \frac{C_r}{C}=\frac{\sum_{i}\langle 0| S_0^zs_{ri}^z|0\rangle}{\sum_{ai}\langle 0| S_0^zs_{ai}^z|0\rangle}.
\label{eq:correlation}
\end{equation}
At the channel-symmetric point $\Gamma = 1$, the overscreened impurity is correlated equally with the two channels, leading to $R_c=1/2$. While for any $\Gamma \neq 1$, the impurity is correlated only with the channel with the stronger coupling, while the other channel decouples. As a result, the value of $R_C$ is expected to go from 0 to 1 around the critical point $\Gamma=1$. Indeed, such a behavior of $R_C$ is confirmed by the NORG calculations, as shown in Fig.~\ref{fig:CorrRatio}(a). The thermodynamic limit behavior can be examined by further studying the derivative of the correlation ratio with respect to $\Gamma$, i.e., $R_C^{\prime}=\partial R_C/{\partial\Gamma}$, which is plotted in Fig.~\ref{fig:CorrRatio}(b). Figure~\ref{fig:CorrRatio}(b) clearly shows that the derivative peaks at the critical point $\Gamma =1$ and this peak becomes more pronounced as the system size $N$ increases. This indicates that $R_C^{\prime}$ tends to diverge at the critical point $\Gamma=1$ as $N \to \infty$, illustrating that the Kondo singlet in 1CK physics is destroyed and the overscreened ground state in 2CK physics emerges when $\Gamma$ tends to 1 from $\Gamma \ne 1$. Therefore, we adopt the spin correlation ratio $R_C$ as an order parameter to characterize the iQPT in the 2CK model driven by the channel asymmetry.

In practical calculations, to characterize the critical behavior around a critical point, we need to do finite-size scaling analysis~\cite{Barber1983,Sandvik2005,Vicari2014,Sorella2016} for an order parameter. \cite{Sachdev2011}It has been known that a diverging characteristic length $\xi$ emerges at the critical point $g_c$ for a quantum phase transition, which scales as $\xi^{-1} \sim |g-g_c|^\nu$ with $\nu$ being a critical exponent and $g$ a control parameter. Meanwhile, an order parameter scales as $|g-g_c|^\beta$ in the vicinity of the critical point, where $\beta$ is a critical exponent. It is emphasized that $\nu$ is uniquely determined by the system Hamiltonian. Here in the 2CK model, the correlation length is taken as the critical crossover scale $\xi^*$ at which the RG flow crosses over from the overscreened 2CK fixed point to the fully screened Fermi liquid fixed point.

\begin{figure}[htp!]
\centering
\includegraphics[width=1.0\columnwidth]{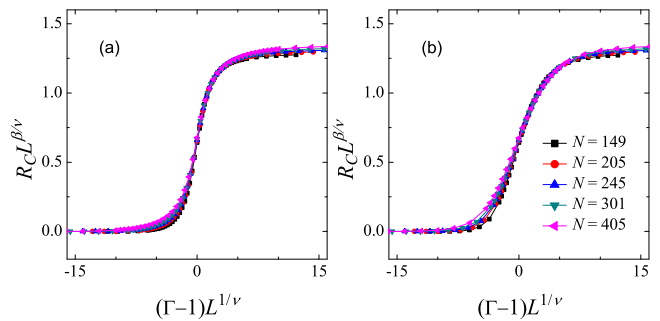}
\caption{\label{fig:Scaling}(color online) Finite-size scaling of the spin correlation ratio $R_C$ with Kondo couplings (a) $J = 1.0$ and (b) $J = 0.5$ in the 2CK model, respectively. The obtained values are $\beta = 0.10 \pm 0.01$ and $\nu = 2.0 \pm 0.1 $.}
\end{figure}

For the spin correlation ratio $R_C$ in the 2CK model, we adopt a standard finite-size scaling form as follows,
\begin{equation}
R_C(L)= L^{-\beta/\nu}F((\Gamma - 1)L^{1/\nu}),
\label{eq:scaling}
\end{equation}
where $F$ is a scaling function and $L$ the number of total sites in all the electronic channels. The order parameter $R_C$ should show a scale-invariant behavior around the critical point $\Gamma = 1$, and thus we plot $R_CL^{\beta/\nu}$ as a function of $(\Gamma - 1)L^{1/\nu}$ in Fig.~\ref{fig:Scaling}(a) and (b) for Kondo couplings $J = 1.0$ and $J = 0.5$, respectively. We explore the values of $\beta$ and $\nu$ such that the curves for different system sizes $N$ collapse on each other. When $\nu=2.0 \pm 0.1$ and $\beta = 0.10 \pm 0.01$, all the curves in Fig.~\ref{fig:Scaling}(a) and (b) respectively collapse to a single one, indicating that a very good data collapse is achieved. The value of critical exponent $\nu = 2$ is in agreement with  the result of CFT and bosonization\cite{Affleck1991,Affleck1995,Fabrizio1995}, as well as with the analysis based on Schmidt gap and negativity\cite{Bayat2014,Alkurtass2016, Bayat2017}. This demonstrates that $R_C$ is an appropriate order parameter for characterizing the iQPT in the 2CK model.

We proceed with considering the 3CK model. At the channel-symmetric point, the 3CK physics is valid with the impurity being overscreened by all the channels. Whereas when one of the Kondo couplings $J_a$ is increased (or decreased), the 3CK fixed point crosses over to a Fermi liquid fixed point with the impurity being screened only by this channel (or to a 2CK fixed point with the impurity being overscreened by the other two channels) on the new temperature scale of $T^* \propto |J_a-J_a^c|^{2.5}$ with $J_a^c$ being the critical coupling at the channel-symmetric point~\cite{Mitchell2014}. Therefore, on fine tuning the Kondo coupling $J_a \to J_a^c$ to the symmetric point, the 3CK physics is obtained with the vanishing temperature scale $T^* \to 0$ and a diverging length scale $\xi^* (\sim 1/T^*) \to \infty$ at the lowest energy scales. On the other hand, similar to the 2CK model, in the channel-asymmetric case, it is expected that there is no correlation between a decoupled channel and the impurity, namely the total correlation in a decoupled channel is 0.

\begin{figure}[htp!]
\centering
\includegraphics[width=0.8\columnwidth]{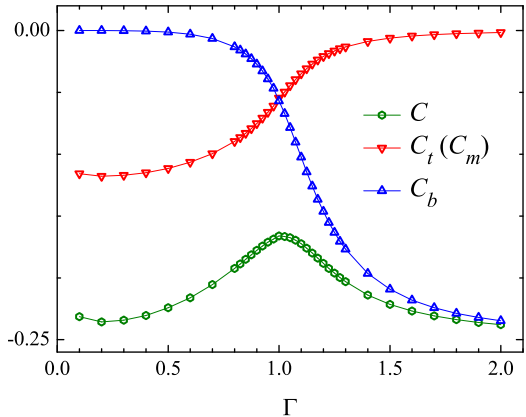}
\caption{\label{fig:Corrs-3CK}(color online) Ground state spin correlations $C$, $C_t (C_m)$ and $C_b$ between the localized impurity and electrons in the electronic channels in the 3CK model, versus $\Gamma$ for system of size $N = 199$ with the fixed coupling $J = 1$.}
\end{figure}

By using the NORG method, we have also studied the 3CK model (Eq.~(\ref{eq:3CKM})) illustrated in Fig.~\ref{fig:Model}(b). Figure~\ref{fig:Corrs-3CK} shows the calculated spin correlation with the fixed coupling $J = 1$. As we see, $C_b \to 0$ when parameter $\Gamma \ll 1$, indicating that the bottom channel decouples while the impurity is correlated equally with the top and middle channels. In contrast, when the parameter $\Gamma \gg 1$, $C_b \to -1/4$ while $C_t$ (or $C_m$) $\to 0$, demonstrating that the impurity is correlated only with the bottom channel while the other two channels decouple. At the channel-symmetric point, the impurity is correlated equally with all the channels, as presented that $C_t$, $C_m$ and $C_b$ cross at the critical point $\Gamma = 1$. On the other hand, for a small parameter $\Gamma \ll 1$, where the impurity is overscreened by the top and middle channels, there is an uptick instead of monotonic decrease towards $-1/4$ for the curves of $C_t$ ($C_m$) and $C$. This results from the nonmonotonic dependence of the Kondo temperature $T_K$ on the Kondo coupling $J_0=J/\Gamma$ for the 2CK model, where $T_K$ depends exponentially on $1/J_0$ and $J_0$ in the two regimes of small and large coupling respectively~\cite{Kolf2007}.

\begin{figure}[htp!]
\centering
\includegraphics[width=1.0\columnwidth]{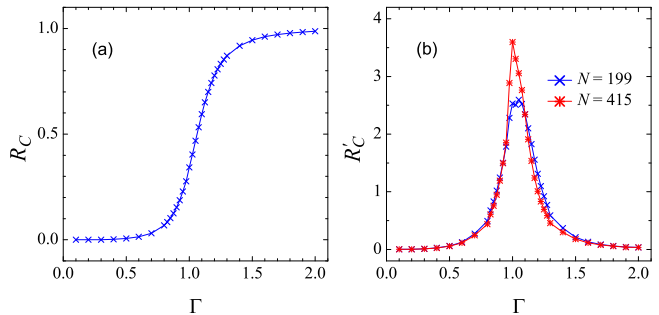}
\caption{\label{fig:CorrRatio3}(color online) (a) Spin correlation ratio $R_C$ as a function of $\Gamma$ for system of size $N = 199$ and (b) derivative of $R_C$ with respect to $\Gamma$, namely $R_C^{\prime}=\partial R_C/{\partial\Gamma}$, for systems of different size $N$ with the fixed coupling $J = 1$ in the 3CK model.}
\end{figure}

In comparison with the case of the 2CK model, definition of the spin correlation ratio $R_C$ is given by
\begin{equation}
R_C= \frac{C_b}{C}=\frac{\sum_{i}\langle 0| S_0^zs_{bi}^z|0\rangle}{\sum_{ai}\langle 0| S_0^zs_{ai}^z|0\rangle}.
\label{eq:correlation-3CK}
\end{equation}
Accordingly, similar to the 2CK model, at the channel-symmetric point $\Gamma = 1$, the overscreened impurity is correlated equally with all the channels, leading to $R_c = 1/3$. On the other hand, for any $\Gamma > 1$, the impurity is correlated only with the bottom channel, while the other two channels decouple. In contrast, for any $\Gamma < 1$, the bottom channel decouples and the impurity is correlated equally with the top and middle channels. As a result, the value of $R_C$ is expected to go from 0 to 1 around the critical point $\Gamma = 1$. The calculated spin correlation ratio $R_C$ and its derivative $R_C^{\prime}=\partial R_C/{\partial\Gamma}$ with respect to $\Gamma$ are presented in Fig.~\ref{fig:CorrRatio3}(a) and (b), respectively. As we see, the value of $R_C$ goes from 0 to 1 around the critical point $\Gamma=1$, and $R_C^{\prime}$ peaks at $\Gamma =1$ with this peak sharpening as the system size $N$ increases, indicating that $R_C^{\prime}$ goes to diverge at the channel-symmetric point $\Gamma=1$ in the thermodynamic limit. Consequently, the behaviors of $R_C$ and $R_C^{\prime}$ in the 3CK model are similar to those in the 2CK model.

\begin{figure}[htp!]
\centering
\includegraphics[width=1.0\columnwidth]{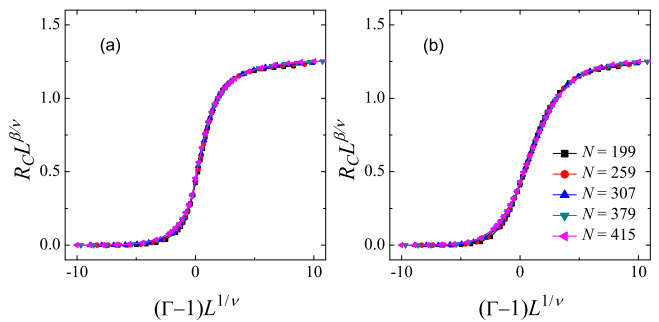}
\caption{\label{fig:Scaling3}(color online) Finite-size scaling of the spin correlation ratio $R_C$ with Kondo couplings (a) $J = 1.0$ and (b) $J = 0.5$ in the 3CK model, respectively. The obtained values are $\beta = 0.10 \pm 0.01 $ and $\nu = 2.5 \pm 0.1 $.}
\end{figure}

Likewise, the same finite-size scaling form of Eq.~(\ref{eq:scaling}) for $R_C$ is used in the 3CK model. Here the correlation length is taken as the critical crossover scale $\xi^*$ at which the RG flow crosses over from the overscreened 3CK fixed point to the fully screened Fermi liquid fixed point ($\Gamma > 1$) or to the overscreened 2CK fixed point ($\Gamma < 1$). The corresponding results are presented in Fig.~\ref{fig:Scaling3}(a) and (b) for fixed Kondo couplings $J = 1.0$ and $J = 0.5$, respectively. We find that the best data collapse is achieved by $\nu=2.5 \pm 0.1 $ and $\beta=0.10 \pm 0.01 $ with $\nu = 2.5$ being in agreement with the result of CFT\cite{Affleck1991,Affleck1995}. Moreover, the values of critical exponent $\beta$ for both 2CK and 3CK models are equal to 0.1, not given in CFT studies. Thus, the spin correlation ratio $R_C$ acts as an appropriate order parameter for characterizing the iQPTs both in the 2CK and 3CK models.

\section{Discussion and summary}
\label{sec:conclusion}
Impurity quantum phase transitions occur in various quantum impurity systems at zero temperature. Multi-channel Kondo system, associated with intermediate-coupling fixed point showing non-Fermi liquid behavior, is a paradigm for exploring iQPTs. According to the CFT, the nontrivial low-temperature properties of an MCK system are determined by the leading irrelevant operator at the overscreened fixed point\cite{Affleck1991}. Specifically, for an MCK model which obeys ${\rm SU}(2)_{\rm spin} \times {\rm SU}(M)_{\rm channel}$ symmetry, it has been convinced that the dimension of the leading irrelevant operator is $1 + \Delta = 1 + 2/(2+M)$, for example, the impurity magnetic susceptibility $\chi_{\text{imp}}\propto T^{2\Delta-1}$ when $T\to 0$ except $M = 2$. On the other hand, at zero temperature, an iQPT can be driven by the channel asymmetry, i.e., the ${\rm SU}(M)_{\rm channel}$ symmetry is broken. More specifically, when one of the couplings is increased, the ground state of the system will transit into a Fermi liquid phase. In contrast, when one of the couplings is decreased, the ground state will behave as that of an ($M-1$)-channel Kondo model with a non-Fermi liquid phase ($M > 2$). For this critical transition, the critical exponent $\nu$ of correlation length is $1/\Delta$ with $\Delta$ being the dimension of the most relevant new operator that appears in the Hamiltonian when the channel symmetry is broken~\cite{Affleck1993,Affleck2005}. Thus for 2CK model $\nu=2$ and for 3CK model $\nu=2.5$. As shown in this work, these generic properties are well manifested by the spin correlation calculations for the 2CK and 3CK models. Especially, the calculated critical exponents $\nu$ are the same as those given by the CFT. Furthermore, the calculations give extra information beyond the CFT, namely the critical exponent $\beta$ of order parameter.

In summary, we introduce an order parameter, namely spin correlation ratio $R_C$, for describing iQPT in multi-channel Kondo system driven by channel asymmetry, based on spin correlation between the impurity and electrons in the electronic channels. By the calculations using the NORG method, we demonstrate that the spin correlation ratio $R_C$ is an appropriate order parameter for describing the iQPTs. By finite-size scaling analysis, the critical exponents $\beta$ of $R_C$ and $\nu$ of correlation length are further determined to be $\beta = 0.10(1),\nu = 2.0(1)$ for the 2CK model and $\beta = 0.10(1),\nu = 2.5(1)$ for the 3CK model, respectively. Moreover, the values of critical exponent $\nu=1/\Delta$ are the same as the results of the conformal field theory for multi-channel Kondo system.

\begin{acknowledgments}

This work was supported by National Natural Science Foundation of China (Grants No. 11934020 and No. 11874421). R. Q. H. was supported by the Fundamental Research Funds for the Central Universities, and the Research Funds of Renmin University of China (Grant No. 18XNLG11). Computational resources were provided by National Supercomputer Center in Guangzhou with Tianhe-2 Supercomputer and Physical Laboratory of High Performance Computing in RUC.

\end{acknowledgments}


\begin{thebibliography}{99}

\bibitem{Kondo1964}J. Kondo, Prog. Theor. Phys. {\bf 32}, 37 (1964).
\bibitem{Hewson1997}A. Hewson, {\it The Kondo Problem to Heavy Fermions} (Cambridge University Press, Cambridge, 1997).
\bibitem{Wilson1975}K. G. Wilson, Rev. Mod. Phys. {\bf 47}, 773 (1975).
\bibitem{Bulla2008}R. Bulla, T. A. Costi, and T. Pruschke, Rev. Mod. Phys. {\bf 80}, 395 (2008).
\bibitem{Blandin1980}P. Nozi\`eres and A. Blandin, J. Phys. France {\bf 41}, 193 (1980).
\bibitem{Andrei1984}N. Andrei and C. Destri, Phys. Rev. Lett. {\bf 52}, 364 (1984).
\bibitem{Tsvelick1984}A. M. Tsvelick and P. B. Wiegmann, Z. Phys. B: Condensed Matter {\bf 54}, 201 (1984).
\bibitem{Tsvelick1985}A. M. Tsvelick, J. Phys. C {\bf 18}, 159 (1985).
\bibitem{Andrei1995}N. Andrei and A. Jerez, Phys. Rev. Lett. {\bf 74}, 4507 (1995)
\bibitem{Andrei2002}G. Zar\'and, T. Costi, A. Jerez, and N. Andrei, Phys. Rev. B {\bf 65}, 134416 (2002).
\bibitem{Affleck1991PRL}I. Affleck and A. W. W. Ludwig, Phys. Rev. Lett. {\bf 67}, 161 (1991).
\bibitem{Affleck1991}I. Affleck and A. W. W. Ludwig, Nucl. Phys. B {\bf 360}, 641 (1991).
\bibitem{Affleck1992}I. Affleck, A. W. W. Ludwig, H. B. Pang, and D. L. Cox, Phys. Rev. B {\bf 45}, 7918 (1992).
\bibitem{Affleck1993}I. Affleck and A. W. W. Ludwig, Phys. Rev. B {\bf 48}, 7297 (1993).
\bibitem{Parcollet1998}O. Parcollet, A. Georges, G. Kotliar, and A. Sengupta, Phys. Rev. B {\bf 58}, 3794 (1998).
\bibitem{Emery1992}V. J. Emery and S. Kivelson, Phys. Rev. B {\bf 46}, 10812 (1992).
\bibitem{Clarke1993}D. G. Clarke, T. Giamarchi, and B. I. Shraiman, Phys. Rev. B {\bf 48}, 7070 (1993).
\bibitem{Pang1991}H. B. Pang and D. L. Cox, Phys. Rev. B {\bf 44}, 9454 (1991).
\bibitem{Sengupta1994}A. M. Sengupta and A. Georges, Phys. Rev. B {\bf 49}, 10020 (1994).
\bibitem{Fabrizio1995}M. Fabrizio, A. O. Gogolin, and P. Nozi\`eres, Phys. Rev. Lett. {\bf 74}, 4503 (1995).
\bibitem{Cichorek2005}T. Cichorek, A. Sanchez, P. Gegenwart, F. Weickert, A. Wojakowski, Z. Henkie, G. Auffermann, S. Paschen,
                      R. Kniep, and F. Steglich, Phys. Rev. Lett. {\bf 94}, 236603 (2005).
\bibitem{Potok2007}R. M. Potok, I. G. Rau, S. Hadas, O. Yuval, and D. Goldhaber-Gordon, Nature (London) {\bf 446}, 167 (2007).
\bibitem{Mebrahtu2013}H. T. Mebrahtu, I. V. Borzenets, H. Zheng, Y. V. Bomze, A. I. Smirnov, S. Florens, H. U. Baranger,
                      and G. Finkel-stein, Nat. Phys. {\bf 9}, 732 (2013).
\bibitem{Keller2015}A. J. Keller, L. Peeters, C. P. Moca, I. Weymann, V. Mahalu, D. Umansky, D. Zar\'and,
                    and G. GoldhaberGordon, Nature (London) {\bf 526}, 237 (2015).
\bibitem{Iftikhar2015}Z. Iftikhar, S. Jezouin, A. Anthore, U. Gennser, and F. Pierre, Nature (London) {\bf 526}, 233 (2015).
\bibitem{Zhu2015}L. J. Zhu, S. H. Nie, P. Xiong, P. Schlottmann, and J. H. Zhao, Nat. Commun. {\bf 7}, 10817 (2015).
\bibitem{Cichorek2016}T. Cichorek, L. Bochenek, M. Schmidt, A. Czulucki, G. Auffermann, R. Kniep, R. Niewa, F. Steglich, and
                      S. Kirchner, Phys. Rev. Lett. {\bf 117}, 106601 (2016).
\bibitem{Mitchell2011}A. K. Mitchell, M. Becker, and R. Bulla, Phys. Rev. B {\bf 84}, 115120 (2011).
\bibitem{Vojta2006}M. Vojta, Philos. Mag. {\bf 86}, 1807 (2006).
\bibitem{Mitchell2014}A. K. Mitchell, M. R. Galpin, S. Wilson-Fletcher, D. E. Logan, and R. Bulla,
                      Phys. Rev. B {\bf 89}, 121105(R) (2014).
\bibitem{Bayat2014}A. Bayat, H. Johannesson, S. Bose, and P. Sodano, Nat. Commun. {\bf 5}, 3784 (2014).
\bibitem{Alkurtass2016}B. Alkurtass, A. Bayat, I. Affleck, S. Bose, H. Johannesson, P. Sodano, E. S. S{\o}rensen,
                       and K. Le Hur, Phys. Rev. B {\bf 93}, 081106(R) (2016).
\bibitem{Bayat2017}A. Bayat, Phys. Rev. Lett. {\bf 118}, 036102 (2017).
\bibitem{He2014}R.-Q. He and Z.-Y. Lu, Phys. Rev. B {\bf 89}, 085108 (2014).
\bibitem{Wang2020}Y. Wang, E. Walter, S. S. B. Lee, K. M. Stadler, J. von Delft, A. Weichselbaum, and G. Kotliar,
                  Phys. Rev. Lett. {\bf 124}, 136406 (2020).
\bibitem{He2015}R.-Q. He, J. Dai, and Z.-Y. Lu, Phys. Rev. B {\bf 91}, 155140 (2015).
\bibitem{Zheng2018}R. Zheng, R.-Q. He, and Z.-Y. Lu, Chin. Phys. Lett. {\bf 35}, 067301 (2018).
\bibitem{Lin2013}C. Lin and A. A. Demkov, Phys. Rev. B {\bf 88}, 035123 (2013).
\bibitem{Barber1983}M. N. Barber, in {\it Phase Transitions and Critical Phenomena}, Vol. {\bf 8} (edited by C. Domb
                    and J. L. Lebowitz), 145--477 (Academic Press, London, 1983).
\bibitem{Sandvik2005}K. S. D. Beach, L. Wang, and A. W. Sandvik, arXiv:cond-mat/0505194.
\bibitem{Vicari2014}M. Campostrini, A. Pelissetto, and E. Vicari, Phys. Rev. B {\bf 89}, 094516 (2014).
\bibitem{Sorella2016}Yuichi Otsuka, Seiji Yunoki, and Sandro Sorella, Phys. Rev. X {\bf 6}, 011029 (2016).
\bibitem{Sachdev2011}S. Sachdev, {\it Quantum Phase Transitions}  (Cambridge University Press, New York, 2011).
\bibitem{Affleck1995}I. Affleck, A. W. W. Ludwig, and B. A. Jones, Phys. Rev. B {\bf 52}, 9528 (1995).
\bibitem{Kolf2007}C. Kolf and J. Kroha, Phys. Rev. B {\bf 75}, 045129 (2007)
\bibitem{Affleck2005}I. Affleck, J. Phys. Soc. Jpn. {\bf 74}, 59 (2005).

\end{thebibliography}

\end{document}